\documentclass[12pt]{iopart}
\usepackage{epsf}
\begin{document}

\newcommand{\ua}{\uparrow}
\newcommand{\da}{\downarrow}
\newcommand{\la}{\langle}
\newcommand{\ra}{\rangle}
\newcommand{\rar}{\rightarrow}
\newcommand{\us}{\uparrow}
\newcommand{\ds}{\downarrow}
\title{Magnetic and superconducting instabilities in the periodic 
Anderson model:
an RPA study}

\author{N. M. R. Peres and M. A. N. Ara\'ujo}

\address{Departamento de F\'{\i}sica, Universidade de \'Evora,
Rua Rom\~ao Ramalho, 59, P-7000-671 \'Evora, Portugal\\
Centro de F\'{\i}sica da Universidade do Minho, Campus Gualtar, 
P-4700-320 Braga, Portugal}

\date{\today}


\begin{abstract}
We study the magnetic and superconducting instabilities of
the periodic Anderson model  with infinite Coulomb repulsion
$U$ in the random phase approximation. 
The N\'eel temperature and the superconducting critical temperature
are obtained as functions of electronic density ({\it chemical pressure})
and  hybridization $V$ ({\it pressure}).
It is found that close to the region where the system exhibits
magnetic order the critical temperature $T_c$ is much smaller
than the N\'eel temperature, in qualitative agreement with some
$T_N/T_c$ ratios found for some heavy-fermion materials. In our study,
 the magnetic and superconducting 
physical behaviour of the system has its origin in the fluctuating
boson fields implementing the infinite on-site Coulomb
repulsion among the $f-$ electrons.    
\end{abstract}
\vspace{0.3cm}
\pacs{74.70.Tx, 75.30.Mb}
\submitto{\JPCM}
\section{Introduction}

The superconducting and magnetic properties of heavy-fermion materials
have recently attracted much attention because of their non-conventional
character. \cite{varma85,estrela00}  These materials have very
large specific heat coefficients $\gamma$, indicating very large effective
quasi-particle masses, hence the designation  {\it heavy fermions}.
Some of these materials, 
order antiferromagnetically at low temperatures 
(examples are  $UAgCu_4$, $UCu_7$, $U_2Zn_{17}$)
 while others (such as $UBe_{13}$, $CeCu_2Si_2$, $UPt_3$) 
order in a superconducting state and others show no ordering (such as
$CeAl_3$, $UAuPt_4$, $CeCu_6$, $UAl_2$).  \cite{varma85}
Some compounds exhibit phases  where  antiferromagnetic
order  coexists  
with unconventional  superconductivity. Examples are: $UPd_2Al_3$ 
($T_N=14.3$ K and $T_c=2$ K),  $UNi_2Al_3$  ($T_N=4.5$ K and $T_c=1.2$ K),
CePd$_2$Si$_2$ ($T_N\sim $ 10 K and $T_c\sim $ 0.5 K)
and CeIn$_3$ ($T_N\sim $ 10 K and $T_c\sim $ 0.15 K).
In the prototype heavy-fermion system Ce$_x$Cu$_2$Si$_2$ the coexistence
of $d-$wave superconductivity  and magnetic order 
was clearly identified in a small range  $x$ values 
around $x\simeq 0.99$.\cite{ishida99}

Systems that exhibit both superconductivity and antiferromagnetism at low
 temperature have 
 ratios between the N\'eel temperature $T_N$
and the superconducting critical temperature $T_c$  of the order of
$T_N/T_c\sim 1-100$.
The coexistence of both types of order can be tuned  by external
parameters such as external pressure 
or changes in the stoichiometry.\cite{ishida99,lonzarich98}

A description of the normal state  properties of the heavy-fermion
systems has been attempted assuming a generalization of the impurity
Anderson model to the lattice. \cite{newns87,Millis87}
In the Anderson lattice the energy of a single 
electron in an $f-$orbital (e. g. $4f^1$)
is $\epsilon_0$, and the energy of two electrons 
in the same $f-$orbital ($4f^2$) is $2\epsilon_0+U$, where $U$ is 
the on-site Coulomb repulsion. The energy of the  $4f^2$ state  is 
much larger than the energy of the $4f^1$ state. Thus, if the charge 
fluctuations  
at the  $f-$orbital are small, the $4f^1$ electron may behave as a 
local moment. 

The complexity of  heavy-fermion  
 systems arises from the interplay between Kondo screening of local moments,
the   antiferromagnetic (RKKY) interaction between the  moments and 
the superconducting correlations between the heavy quasi-particles. 
The local moments form in  partially filled $f$ shells of Ce and U ions.
The absence of magnetic order in some cases could perharps be due to 
complete Kondo screening (below the Kondo temperature $T_K$)
or to  a spin liquid  arrangement of the local moments. 
In the normal non-magnetic state the Anderson lattice model predicts 
Fermi-liquid like behaviour and explains the main features at
low temperatures, such as the  large effective masses and  the Kondo
resonance near the Fermi level.  But the main technical difficulty 
is the competition between the Kondo compensation of the localized spins
and the magnetic interaction between them. This interaction
is mediated by the conduction electrons ($RKKY$-type). Related to
this competition is the effectiveness of the compensating cloud around each 
$f$-site. The size of this cloud has been subject of controversy. While some
arguments show that it should be a large scale of the order of 
$v_F/T_K$ \cite{exhaust1},  other arguments claim to be $\sim a$
($a$ is the lattice constant). \cite{exhaust2}
This is a relevant issue and is related to Nozi\`{e}res exhaustion
problem which states that there are not enough conduction electrons to screen
the $f$-moments.

It has been proposed that the mechanism for superconductivity lies in  
the strong
Coulomb interaction between the $f$-electrons, not in a  phonon mediated
 attraction. Using Coleman's \cite{coleman84} slave boson
formalism together with a large-$N$ approach, various attempts have been
made to search for the existence of an effective interaction which might be
responsible for superconductivity in the 
infinite-$U$ Anderson-lattice model. It was proposed
\cite{lavagna87} that  slave boson fluctuations  can provide an effective 
attraction  between the electrons to leading order in $1/N$.
Later, a calculation of the electron-electron  scattering amplitude to order
 $1/N^2$  revealed an effective attractive interaction in the $p$ and $d$ 
channels,
 which was interpreted as a manifestation of the $RKKY$ interaction, showing 
that spin fluctuations are an important mechanism.
\cite{houghton88}
The inclusion of  $f^0$, $f^1$ and $f^2$ states,
using two sets of slave bosons, was also considered
in the context of the Anderson lattice as  a possible description
of high-$T_c$ superconductors.\cite{rasolt87}

The magnetic order in the ground state of Kondo Insulators
 has been studied by Dorin and 
Schlottmann in the framework of the Anderson-lattice model\cite{dorin92}. 
The same authors have later  studied the effect
of orbital degeneracy and finite $U$ on 
a ferromagnetic ground state (their approach 
 did not generate RKKY interactions, thus preventing
the study of antiferromagnetic order).\cite{dorin93}

In this work we consider the slave-boson approach to 
the infinite-$U$ Anderson-lattice model. We treat the boson
fields at the mean-field level,  thereby enforcing the constrain
of one $f$-electron (at most) per site  only on the average. 
By splitting the boson operator into 
a condensate part   and an above-the-condensate term, which describes
fluctuations, we compute the magnetic and pairing
susceptibilities at the random-phase approximation (RPA) level.
For spin 1/2 particles  the condensate density at moderate
temperatures does not change much relative to its ground-state
value. Therefore we do not expect our results to be of less quality
then those characterizing  the ground-state properties. 
We search for 
the critical temperatures ($T_N$ and $T_c$) at which anti-ferromagnetic 
order or
superconducting ($s$ or $d$-wave) pairing occurs in a  normal 
non-magnetic system.  
We find  that the value of  $T_c$  is much smaller than the   
magnetic temperature $T_N$. 
Unlike $T_N$, the superconducting temperature
monotonically increases with externally applied pressure.

\section{The model and the RPA solution}

The PAM Hamiltonian is given by

\begin{equation}
H=H^0_c+H_f^0+H_{cf}+H_U\,,
\label{hi}	
\end{equation}
where
\begin{eqnarray}
H_f^0&=&\sum_{i,\sigma}(\epsilon_0-\mu)f_{i,\sigma}^{\dag}f_{i,\sigma}\,,\\
H_c^0&=&\sum_{\vec k,\sigma}(\epsilon_{\vec k}-\mu)
c_{\vec k,\sigma}^{\dag}c_{\vec k,\sigma}\,,\\
H_{cf}&=&V\sum_{i,\sigma}\left(c_{i,\sigma}^{\dag}f_{i,\sigma}
+f_{i,\sigma}^{\dag}c_{i,\sigma}\right)
\label{hcf}\,,\\
H_U&=&U\sum_{i,}n_{i,\us} n_{i,\ds}\,.
\end{eqnarray}
The $c$ and $f$ operators  are fermionic and obey the usual anti-commutation
relations. The hybridization potential $V$ is assumed to be momentum
independent. The term $H_U$ represents the strong on-site repulsion 
between the $f$-orbitals. We 
 consider  $U=\infty$. We implement the condition $U=\infty$  
within the slave-boson formulation due 
to Coleman, \cite{coleman84} in which the  empty $f$-site is represented
by a slave boson $b_i$ and  the physical operator $f_i$ in equation  
(\ref{hcf}) is replaced with  ${b^{\dagger}}_i f_i$ together
with the constrains of only one $f-$electron per site. The implementation of
this constraint amounts to introducing 
 a Lagrange multiplier $\lambda$ which will renormalize the
bare $f-$level energy from $\epsilon_0$ to 
$\epsilon_f=\epsilon_0+\lambda$.
We split the boson operators in two terms
\begin{equation}
b^{\dagger}_{\vec q}=\sqrt{N}\sqrt{z}\delta_{0,\vec q}+B^\dagger_{\vec q}\,,
\end{equation}
where $z$ represents the boson condensate and $B^\dagger_{\vec q}$
represents the fluctuations above the condensate.
This procedure leads in leading order
to a mean field Hamiltonian \cite{mana00,nuno01}.
 The corresponding mean field equations can be written
in terms of the Fourier transform of the Green's functions 
	\begin{eqnarray}
	G_{ff,\sigma}(\vec k,\tau-\tau') & = & 
	-\la T_{\tau}f_{\vec k,\sigma}(\tau)
	f^{\dag}_{\vec k,\sigma}(\tau')\ra\,,\\
	G_{cc,\sigma}(\vec k,\tau-\tau') & = & 
	-\la T_{\tau}c_{\vec k,\sigma}(\tau)
	c^{\dag}_{\vec k,\sigma}(\tau')\ra\,,\\
	G_{cf,\sigma}(\vec k,\tau-\tau') & = & 
	-\la T_{\tau}c_{\vec k,\sigma}(\tau)
	f^{\dag}_{\vec k,\sigma}(\tau')\ra\,,
	\end{eqnarray}
as

	\begin{equation}
	z  =  1- \frac{T}{N_s} \sum_{\vec{k},\sigma} \sum_{i\omega_n}
 	{\cal G}_{ff,\sigma} (\vec{k},i\omega_n)
\label{z}\,,
	\end{equation}
and 
	\begin{equation}
	\epsilon_f   =\epsilon_0  - \frac{VT}{\sqrt{z}N_s}
 	\sum_{\vec{k}, \sigma} \sum_{i
	\omega_n} G_{cf,\sigma} (\vec{k}, i\omega_n)\,,
	\end{equation}
where $N_s$ denotes the number of lattice sites.
Equation (\ref{z}) states that the  mean number of 
electrons at an $f$-site is $n_f=1-z$.
For a given number of particles per site, $n$, these
equations  must be supplemented with the  particle conservation condition 
which yields the chemical potential $\mu$ for any temperature:
	\begin{equation}
	n =  1-z+\frac{T}{N_s} \sum_{\vec{k},\sigma} \sum_{i\omega_n}
 	G_{cc,\sigma} (\vec{k},i\omega_n)\,.\\
\label{mean}
	\end{equation}

The fluctuations beyond the mean field  approach are described by the 
Hamiltonian
	\begin{equation}
	H_{fluct}=\frac {V}{\sqrt N}\sum_{\vec k,\vec q,\sigma}
	(c^\dag_{\vec k,\sigma}f_{\vec q,\sigma}B^\dag_{\vec k-\vec q}+
	B_{\vec k-\vec q}f^\dag_{\vec q,\sigma}c_{\vec k,\sigma})\,,
	\end{equation}
and will be considered in the calculation of 
the magnetic susceptibility and  
superconducting correlation functions below.
The   calculation, even at the RPA level, of the correlation functions
requires  the knowledge of the  boson propagator. 
The full calculation of the latter is
a technically difficult problem  by itself, and is still unsolved. 
There are, however, $1/N$ calculations of $D(\vec k,\tau-\tau')$.
\cite{Millis87,lavagna87,houghton88}
Here we follow the 
work  of Evans \cite {evans1} and use an asymptotic form for the boson
propagator given by
	\begin{equation}
	D(\vec k,\tau-\tau')=\la T_{\tau}B_{\vec k,\sigma}(\tau)
	B^{\dag}_{\vec k,\sigma}(\tau')\ra\sim\frac {1}{\lambda}\,.
\end{equation}
We also adopt the same approximation
for the propagator ${\bar  D}(\vec k,\tau-\tau')=
\la T_{\tau}B_{\vec k,\sigma}^\dag(\tau)B_{\vec k,\sigma}(\tau')\ra$.
In the calculation  below we shall use  mean field fermionic propagators.

The transverse spin susceptibility for the $f$ electrons is defined as
\begin{equation}
\chi_{-+}(\vec q,i\omega_n)=\mu^2_B\int_0^\beta d\,\tau e^{i\omega_n\tau}
\la T_\tau S^-(\vec q,\tau)S^+(\vec q,0)\ra\,,
\end{equation}
where $\beta=1/T$  is the inverse temperature,
$T_\tau$ is the chronological order operator (in imaginary time),
$S^-(\vec q)= \sum_{\vec p}f^\dag_{\vec p,\da}f_{\vec p+\vec q,\ua}$
and $S^+(\vec q)=[S^-(\vec q)]^\dag$. The calculation at the
RPA level yields
	\begin{equation}
	\chi_{+,-}^f(\vec q,i\omega_n)=\frac {\bar\Gamma^{ff}_{ff}
         (\vec q,i\omega_n)
         [1-J\bar\Gamma^{cf}_{fc}(\vec q,i\omega_n)]}
         {[1-J\bar\Gamma^{cf}_{fc}(\vec q,i\omega)]^2             
	-J^2\bar\Gamma^{ff}_{ff}(\vec q,i\omega_n)\bar
	\Gamma^{cc}_{cc}(\vec q,i\omega_n)}\,,
\label{chi}	
	\end{equation}
where $J=V^2/(N\lambda)$. The result (\ref{chi}) holds
for all values of $n_f$ and is a generalization
of that obtained by Evans \cite{evans1,evans2} for the case $n_f=1$.
The functions $\bar\Gamma(\vec q,i\omega_n)$ above are given by:
	\begin{eqnarray*}
	\bar\Gamma^{ff}_{ff}(\vec q,i\omega_n)&=&-\frac 1 {\beta}\sum_{\vec p,i
        \omega_m} G_{ff}(\vec p,i\omega_m)G_{ff}
	(\vec p+\vec q,i\omega_m+i\omega_n)
	\,,\\
	\bar\Gamma^{cc}_{cc}(\vec q,i\omega_n)&=&-\frac 1 {\beta}\sum_{\vec p,i
        \omega_m} G_{cc}(\vec p,i\omega_m)G_{cc}
        (\vec p+\vec q,i\omega_m+i\omega_n)
	\,,\\
	\bar\Gamma^{cf}_{fc}(\vec q,i\omega_n)&=&-\frac 1 {\beta}\sum_{\vec p,i
        \omega_m} G_{cf}(\vec p,i\omega_m)
	G_{fc}(\vec p+\vec q,i\omega_m+i\omega_n)
	\end{eqnarray*}
There are three possible superconducting pairing susceptibilities that one can
define. These refer to Cooper pairs of either $c-$electrons 
or $f-$electrons,
and a hybrid Cooper pair with a $c-$ and an $f-$electron. We consider the
correlation function:
	\begin{equation}
	\Delta_{dd}(\vec q,i\omega_n)=\int_0^\beta e^{i\omega_n\tau}
	\sum_{\vec k_1,\vec k_2}\eta(\vec k_1)\eta(\vec k_2)
	\la T_{\tau}d_{\vec k_1,\ds}(\tau)d_{-\vec k_1+\vec q,\us}(\tau)
        d_{\vec k_2,\ds}^{\dag}d^\dag_{-\vec k_2+\vec q,\us} \ra\,,
\label{del}	
	\end{equation}
where $d=c,f$ and $\eta(\vec k)$ is the 
Cooper pair structure factor, assumed to be either extended $s-$wave
or $d-$wave.  The hybrid pairing correlation function is 
defined as
	\begin{equation}
	\Delta_{cf}(\vec q,i\omega_n)=\int_0^\beta e^{i\omega_n\tau}
	\sum_{\vec k_1,\vec k_2}
	\la T_{\tau}f_{\vec k_1,\ds}(\tau)c_{-\vec k_1+\vec q,\us}(\tau)
        c_{\vec k_2,\ds}^{\dag}f^\dag_{-\vec k_2+\vec q,\us} \ra\,.
\label{delcf}		
	\end{equation}
This definition has been used previously in a mean field study of the
Kondo lattice \cite{gusmao}. At the RPA level the Cooper pair correlation
function (\ref{del}) is given by
	\begin{equation}
	\Delta_{dd}(\vec q,i\omega_n)= \Gamma^{dd}_{dd}(\vec q,i\omega_n)+
	\frac {J\Gamma^{fc}_{cc}(\vec q,i\omega_n)
        \Gamma^{cc}_{fc}(\vec q,i\omega_n)}{1-
        J[\Gamma^{fc}_{cf}(\vec q,i\omega_n)+
	 \Gamma^{ff}_{cc}(\vec q,i\omega_n)]}\,,	
\label{delrpa}
	\end{equation}
and the function  (\ref{delcf}) is given by
	\begin{equation}
	\Delta_{cf}(\vec q,i\omega_n)= \Gamma^{ff}_{cc}(\vec q,i\omega_n)+
	\frac {J\Gamma^{fc}_{fc}(\vec q,i\omega_n)
        \Gamma^{ff}_{cc}(\vec q,i\omega_n)}
	{[1-J\Gamma^{fc}_{fc}(\vec q,i\omega_n)]^2
        +J^2[\Gamma^{ff}_{cc}(\vec q,i\omega_n)]^2}\,.	
\label{delcfrpa}	
	\end{equation}
The  $\Gamma(\vec q,i\omega_n)$ functions appearing in the
previous expressions are given by
	\begin{eqnarray*}
	\Gamma^{cc}_{cc}(\vec q,i\omega_n)&=&\frac 1 {\beta}\sum_{\vec p,i
        \omega_m}\eta^2(\vec p) 
	G_{cc}(\vec p,i\omega_m)G_{cc}(-\vec p+\vec q,-i\omega_m+i\omega_n)
	\,,\\
	\Gamma^{cc}_{fc}(\vec q,i\omega_n)&=&\frac 1 {\beta}\sum_{\vec p,i
        \omega_m} \eta(\vec p)
	G_{cc}(\vec p,i\omega_m)G_{fc}(-\vec p+\vec q,-i\omega_m+i\omega_n)
	\,,\\
	\Gamma^{ff}_{cc}(\vec q,i\omega_n)&=&\frac 1 {\beta}\sum_{\vec p,i
        \omega_m} G_{ff}(\vec p,i\omega_m)
	G_{cc}(-\vec p+\vec q,-i\omega_m+i\omega_n)\,,\\
	\Gamma^{fc}_{cf}(\vec q,i\omega_n)&=&\frac 1 {\beta}\sum_{\vec p,i
        \omega_m} G_{fc}(\vec p,i\omega_m)
	G_{cf}(-\vec p+\vec q,-i\omega_m+i\omega_n)\,,
	\end{eqnarray*}
\section{Superconducting and magnetic instabilities}

The magnetic and superconducting instabilities of the system are signaled
by the poles of the corresponding susceptibilities. 
Therefore, we search for the temperature $T$ at which the 
denominators in the RPA expressions for the susceptibilities  vanish:
	\begin{eqnarray}
	K_m(\vec Q,0)&=&[1-J\bar\Gamma^{cf}_{fc}(\vec Q,0)]^2             
	-J^2\bar\Gamma^{ff}_{ff}(\vec Q,0)\bar
	\Gamma^{cc}_{cc}(\vec Q,0)\label{km}\,,\\
	K_{dd}(0,0)&=&1-
        J[\Gamma^{fc}_{cf}(0,0)+
	\Gamma^{ff}_{cc}(0,0)]\label{kd}\,,\\
	K_{cf}(0,0)&=&[1-J\Gamma^{fc}_{fc}(0,0)]^2
        +J^2[\Gamma^{ff}_{cc}(0,0)]^2\,,
	\label{kcf}
\end{eqnarray}
where $\vec Q=(\pi,\pi,\pi)$ and $K_m(\vec Q,0)$, $K_{dd}(0,0)$, and
$K_{cf}(0,0)$ are the Stoner factors of the correlation
functions (\ref{chi}), (\ref{delrpa}), and (\ref{delcfrpa}), respectively.
Since heavy-fermion
materials are antiferromagnetic materials we seek for poles of 
$K_m(\vec Q,0)$ at the antiferromagnetic wave vector $\vec Q$. From
the definitions of the $\Gamma(\vec q,i\omega_n)$ functions
we see that the Cooper pair structure factor $\eta(\vec p)$ does not 
appear in the Stoner factors $K_{dd}(0,0)$ and
$K_{cf}(0,0)$. Moreover, we shall see below that   the
solutions to $K_{dd}(0,0)=0$ and
$K_{cf}(0,0)=0$ both lead to the same critical temperature. Therefore, the
 system's tendency  for a certain Cooper pair symmetry only shows
up in the intensity of $\Delta_{dd}(0,0)$ or $\Delta_{cf}(0,0)$, which
is controlled by the numerator of these functions. We also see that
both antiferromagnetism and superconductivity are controlled
by the same interaction parameter $J$, which, in turn, depends on 
hybridization only.

\begin{figure}
\begin{center}
\epsfxsize=10cm
\epsfbox{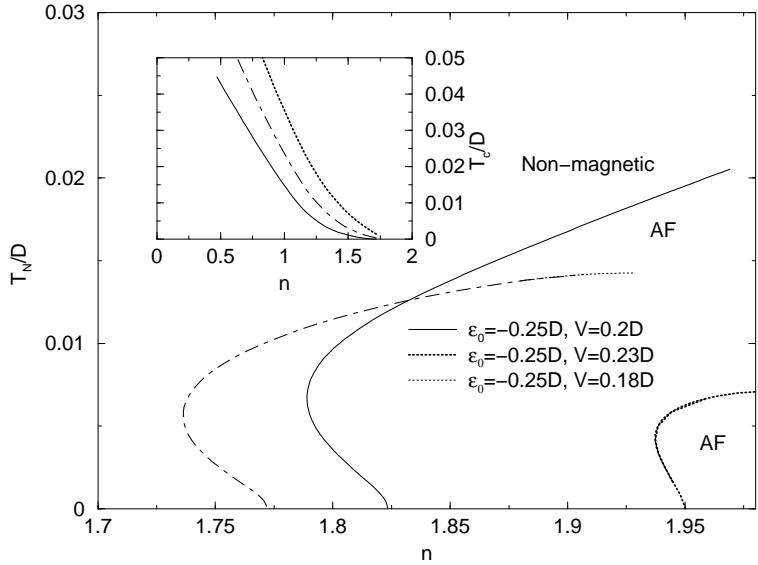}
\end{center}
\caption{N\'eel Temperature as functions of the electronic density
$n$. The inset shows the superconducting critical temperature
as function of $n$. The temperatures are normalized by half bandwidth, $D=6t$
and $t=1$.}
\label{tn}
\end{figure}

In Figure \ref{tn} we show 
a plot of the N\'eel and superconducting temperatures
as functions of the total electronic density $n$. It is seen that 
antiferromagnetism can only occur in a very small region of
electronic density. Furthermore, the increase of $T_N$ when $n\rar 2$
corresponds to an increase of the density of $n_f$ electrons
towards the Kondo limit ($n_f=1$). It is also clear that
$T_N$ is not a monotonically  increasing function of $J$. 
Upon reducing $V$, a larger
range of electronic densities can be reached  where
antiferromagnetic order can be found. From the inset of Figure \ref{tn}
we see that the superconducting temperature $T_c$ is very small 
(about $T_c\sim T_N/50$)
close to the density region where the system exhibits antiferromagnetic order.

\begin{figure}
\begin{center}
\epsfxsize=12cm
\epsfbox{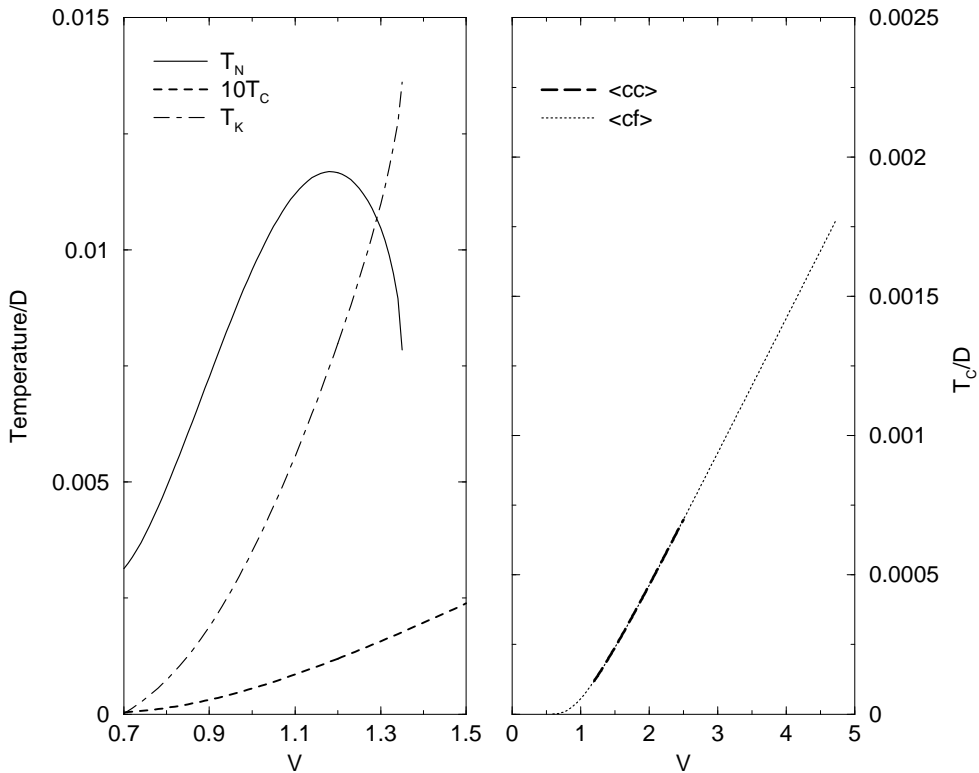}
\end{center}
\caption{{\bf Left: }N\'eel Temperature $T_N$ and superconducting 
critical temperature
$T_c$ as function of $V$, for a constant ratio
of $V/t=1.2$,  electronic density $n=1.8$, and
$\epsilon_0=-0.25D$ . {\bf Right: } $T_c$ over a very large (nonphysical)
values of $V$. Note that $T_c\ll T_N$ always.
$\la cc\ra$ and  $\la cf \ra$ means that $T_c$ has been computed using
equations \ref{kd} and \ref{kcf}, respectively. Both give the
same results. }
\label{pressure}
\end{figure}

The dependence of
the N\'eel  and  superconducting temperatures on pressure has been measured 
in some heavy-fermion systems \cite{lonzarich98,groche96,kawasaki2001}. 
In those studies
the N\'eel temperature is found to decrease as the applied  pressure increases 
and  superconducting  order is found to develop in a limited range of applied
pressures, when the N\'eel temperature is reduced below $\sim$1K. 
Let us now see how the critical temperatures in our model
vary with
the model parameters which, in principle, should depend on externally applied
pressure. Increasing  pressure should, presumably, make both
the hybridization $V$ and the conduction band hopping $t$ increase
\cite{lacroix99,mana02}. In Figure \ref{pressure} we present
 $T_N$ and $T_c$ versus $V$, taking the ration $V/t$ constant. We
see that above a certain value of $V$ the magnetic order disappears but 
the superconducting order remains. We also find that
close to the region where
the magnetic order vanishes,  $T_c$ is much smaller than the maximum
value attained by $T_N$. This is in qualitative agreement with experimental
data on some Cerium compounds (e.g. CeIn$_3$), 
where the ratio  $T_N/T_c\sim 100$. Other examples
are: CeCu$_2$(Si$_{1-x}$Ge$_x$)$_2$, where $T_c$ as function of pressure 
displays  a positive curvature; and  CeRhIn$_5$, where the  $T_c$ curve 
is almost parallel to pressure axis. \cite{kitaoka02}
 For CeCu$_2$Ge$_2$ and CeCu$_2$Si$_2$ the 
$T_c$ curve initally stays almost parallel to the pressure axis, but it shoots
up  above a certain pressure.\cite{kitaoka00}
Although $T_c$ keeps increasing as $V$ increases, it never reaches
values comparable with the maximum value of $T_N$, even for
unreasonable values of $V$ as we can see in the right panel of 
Figure \ref{pressure}. 
We believe that a better treatment of the boson
propagator will lead to a decrease of $T_c$ in agreement with the experiments.
We remark that the above  calculation of $T_c$ is only valid in the
in a situation where the system is
non-magnetic  because we have not calculated 
$K_{dd}(0,0)$ or $K_{cf}(0,0)$ in the magnetically ordered phase.
Furthermore, when $T_c$ is  small, the approximation employed for the 
boson propagator should be improved by including its low energy
part. 

For comparison 
we also plot  the temperature $T_K$ defined as the difference between the 
renormalized $f-$level energy $\epsilon_f$  and the chemical potential
\cite{coleman84,lavagna87,houghton88}. 
This  can be very 
different from the lattice Kondo temperature \cite{iglesias97} 
in the non-magnetic system. Nevertheless, the combined behavior
of $T_N$ and $T_K$ represents  the well-known Doniach diagram showing
the interplay between RKKY and Kondo screening effect. For small values
of $V$, $T_K$ is exponentially small and the system shows
antiferromagnetic order. On the other hand, as $V$ increases the Kondo
temperature grows, leading to Kondo compensation of the $f-$moments
and to a  decreasing N\'eel temperature. For even larger
values of $V$  complete disappearance
of the magnetic order takes place and the system 
shows paramagnetic behavior (assuming there are enough conduction
electrons to compensate all the $f-$local moments). 
We have also computed the superconducting critical temperature
from both equations (\ref{kd}) and ({\ref{kcf}}) and  obtained
the same $T_c$, as can be seen in the right panel of 
Figure \ref{pressure}. Along the $T_N$ curve, $n_f$ decreases from 1
to 0.8, as $V$ increases, and  $n_f \approx 0.85$  when
$T_N$ is maximum.

Both Figure \ref{tn} and Figure \ref{pressure} show  similar
behaviour near the point where $T_N\rar 0$. In both cases $T_c$ starts
to increase with a positive curvature. Although in many heavy-fermion
systems $T_c$ presents a negative curvature, there are examples
where a positive curvature have been observed, such as 
CePd$_2$Si$_2$ under chemical  pressure 
($T_N=10$ K, $T_c=0.2$ K, therefore $T_N/T_c=50$)
\cite{mathur00}.

Since our treatment does not take  competition
between magnetism and superconductivity into account,
 we cannot tell whether  finite values of $T_c$ and
$T_N$ imply that both types of order will be present at low temperature.
Nevertheless, we found in previous work
\cite{mana02},  at the simplest mean field
level, that magnetism and superconductivity may
coexist in the system. It follows  from the above remarks
that the calculation
of $T_c$ when $T_N$ is finite requires both the introduction of
a better approximation for the boson propagator and extra 
electronic propagators describing the  antiferromagnetic order
in the system, as was done in the description of spin waves in the
magnetically ordered Mott insulator.\cite{spinwave} 
\ack

The authors want to thank P. D. Sacramento for 
comments on the manuscript. 


\end{document}